\documentclass[doublecol]{epl2}
\usepackage{graphics,epsfig}
\usepackage{dcolumn}
\usepackage{bm}
\usepackage{color}
\usepackage{subfig}
\newcommand{\PreserveBackslash}[1]{\let\temp=\\#1\let\\=\temp}
\newcolumntype{C}[1]{>{\PreserveBackslash\centering}p{#1}}
\newcolumntype{R}[1]{>{\PreserveBackslash\raggedleft}p{#1}}
\newcolumntype{L}[1]{>{\PreserveBackslash\raggedright}p{#1}}

\title{Influence of Reciprocal links in Social Networks}
\shorttitle{Influence of Reciprocal links in Social Networks}
\author
{
    Yu-Xiao Zhu\inst{1,2} \and
    Xiao-Guang Zhang\inst{3} \and
    Gui-Quan Sun\inst{3,4} \and
    Ming Tang\inst{1} \and
    Tao Zhou\inst{1} \and
    Zi-Ke Zhang\inst{4}\footnote{\email{Corresponding author: zhangzike@gmail.com}}
}
\shortauthor{Y.-X. Zhu \etal}

\institute
{
  \inst{1} Web Sciences Center, University of Electronic Science and Technology of China, Chengdu 610051, P. R. China\\
  \inst{2} Beijing Computational Science Research Center, Beijing 100084, P. R. China\\
  \inst{3} Department of Mathematics, North University of China, Taiyuan 030051, P. R. China\\
  \inst{4} Institute of Information Economy, Hangzhou Normal University, Hangzhou 310018, P. R. China\\
}

\pacs{89.20.Ff}{Computer science and technology}
\pacs{89.75.Hc}{Networks and genealogical trees}
\pacs{89.65.-s}{Social and economic systems}

\abstract{In this Letter, we empirically study the influence of reciprocal links, in order to understand its role in affecting the structure and function of directed social networks. Experimental results on two representative datesets, \emph{Sina Weibo} and \emph{Douban}, demonstrate that the reciprocal links indeed play a more important role than non-reciprocal ones in both spreading information and maintaining the network robustness. In particular, the information spreading process can be significantly enhanced by considering the reciprocal effect. In addition, reciprocal links are largely responsible for the connectivity and efficiency of directed networks. This work may shed some light on the in-depth understanding and application of the reciprocal effect in directed online social networks.}

\begin{document}

\maketitle{}

\section{Introduction}

Nowadays, the emergence of social networks and affiliated applications have triggered an increasing attention from various disciplines, ranging from
studying the social interactions and spreading patterns in social sciences \cite{Jackson02, Kwak10} to uncovering the underlying structure and dynamics in mathematics and physics \cite{Palla07,Castellano09}.
Generally, social networks can be classified into two typical classes according to the edge properties: undirected and directed.
Undirected social networks, such as \emph{Flick} \cite{websites} and \emph{Okut} \cite{websites}, do not allow two users to be connected unless the relation is mutually confirmed, hence,
they are normally regarded as equivalent individuals in graph theory. Comparatively, directed social networks, such as \emph{Twitter} \cite{websites} and \emph{Epinions} \cite{websites}, contain both \emph{unidirectional} and \emph{bidirectional} links, which
consequently build up a so-called \emph{follower/followee} structure \cite{HubermanB2009, Gayo2011, Grabowicz2012}. An online user is
considered as a follower once s/he collects some other users as friends (followees), and puts close attention to them via automatically receiving
their real-time information, as well as online activities \cite{Cimini2012}. A considerable fraction of those followees would also give positive feedback
and add some of their followers with similar interests as online neighbors. Subsequently, such intermediate directed structure
property, namely \emph{reciprocity} \cite{Falk2006}, facilitates a great deal of attention from the scientific community. Nowak and Sigmund discussed that the indirect reciprocity would
help in building reputation systems, judging morality and eventually promote the cooperation level \cite{Rong2010} and benefit the evolution of natural selection \cite{Nowak2005} in both social environment \cite{Rockenbach2006, West2007} and supply networks \cite{Ge2012}. Pereira \emph{et al.} experimentally discussed that negative reciprocity, because of lower cost and less effort, was somehow more favored than the positive reciprocity \cite{Pereira2006}.
Moreover, the power of reciprocity \cite{Diekmann2004} does not only play a vital role in social economic systems \cite{Berg1995,Fehr2002} and human social organizations \cite{Milinski2002, Fehr2003}, but also has been found wide applications in characterizing the property \cite{Diego2004, Zlatic2006}, maintaining the structure \cite{ZhouT2011, Lv2011}, and uncovering the underlying function of directed social networks \cite{Matus2009, Cui2012}.

Typically, the simplest definition of reciprocity, $r$, can be quantified
as the ratio of the number of bidirectional links, $L^{\leftrightarrow}$, to the total number of links $L$ \cite{Newman2002,Serrano2003},
\begin{equation}
r=\frac{L^{\leftrightarrow}}{L}.\label{eq_r}
\end{equation}
For the extreme cases, $r=0$ represents an absolute directed network where all links are unidirectional, and $r=1$ stands for a complete
undirected network where all links are reciprocal. Therefore, the value of $r$ measures the probability that two nodes of a given link are mutually connected.
However, Garlaschelli and Loffredo \cite{Diego2004} argued that Eq. (\ref{eq_r}) failed to precisely describe the full network information,
For example, the network density and self-loops can significantly affect the final measurement of mutual connections.
Alternatively, they proposed a new measure of reciprocity considering the ordering of different networks according to
their actual degree of reciprocity, denoted as
\begin{equation}
\rho=\frac{\frac{L^{\leftrightarrow}}{L}-\bar{a}}{1-\bar{a}}=\frac{r-\bar{a}}{1-\bar{a}},
\label{rr}
\end{equation}
where $\bar{a}=L/N(N-1)$ measures the ratio of observed links to all possible directed links (namely link density). Based on this improved measure, Zlati\'{c} \emph{et al.} \cite{Zlatic2006}
reported that the reciprocity of Wikipedia \cite{websites} could be very similar to other directed networks,
but having a stronger reciprocity than the networks of associations and dictionary terms, and smaller than that of World Wide Web.
Besides that, they found that such a measure is quite stable for different scales of Wikipedia networks, hence is very important for
describing the structure and evolution of wiki-based networks. Bogu{\~n}{\'a} \emph{et al.} \cite{Serrano2005} found that reciprocal connections
played a crucial role in constructing the giant connected component and possibly affecting the Web navigability. Futhermore, Serrano \emph{et al.} \cite{Serrano2007}
provided an in-depth study of the effect of reciprocal links on degree-degree correlations and clustering. They
found that reciprocal links indeed organized the local subgraphs of the World Wide Web network by forming start-like structures,
as well as cliques and communities, which contained highly interconnected pages.
What's more, Gorka \emph{et al.} \cite{Gorka2008} argued that the reciprocity was largely dependent on degree-degree correlation, which, consequently could partially reveal
the underlying hierarchical structure of networks.
Zlati\'{c} and \v{S}tefan\v{c}i\'{c} \cite{Zlatic2009} discussed the influence of reciprocity on vertex degree distribution
and degree correlations. They found that networks driven by reciprocal mechanisms are significantly different from static networks.

In this Letter, we shall provide a specific empirical study of the reciprocity influence on the structure and function of social networks.
In particular, we apply a widely used epidemic spreading model
\cite{Pastor2001, Barthelemy2004, Perra2012} to observe the effect of reciprocity on information spreading. Numerical results show that reciprocal
links can noticeably enhance the speed of information spreading. In addition, we show how reciprocal links affect the structure robustness as percolation catalysts in maintaining the global connectivity by investigating the avalanche of giant components, the network susceptibility and the network distance \cite{Menczer2004, ChengXQ2010JSM}.

\section{Data and Analysis}

In this Letter, we consider two representative directed social networks:
(i) \emph{Sina Weibo}~\cite{sina_site}: the largest Chinese microblogging website, where a user (\emph{follower}) can add
others as his/her friends (\emph{followee}) and automatically receive their posts and events. In addition, users
can forward, comment or share their followees' news on their own post walls; (ii) \emph{Douban}~\cite{douban_site}:
the largest Chinese website for reviewing online movies, books, and music. Besides users' generally proactive contribution, \emph{Douban} also provides services via its recommendation mechanism, which can suggest items of users' potential interests by mining their personalized preferences. Similar with \emph{Sina Weibo}, users in \emph{Douban} can also build follower-followee relationship with each other.
\begin{table}[htbp]
\caption{Basic statistics of the two observed data sets.
$N=|V|$ and $M=|E|$ are the total number of nodes and links,
respectively, $\rho$ is the network reciprocity denoted by Eq. (\ref{rr}), and $S=M/N(N-1)$ denotes the network sparsity.}
\begin{tabular}{l r r r r}
\hline\hline
 Data sets      &    $N$    &    $M$    & $\rho$    &    $S$     \\
\hline
\emph{Sina Weibo}     &    3,592   &    12,522  &   0.307      &  9.7$\times10^{-4}$   \\
\emph{Douban}         &    3,097   &    6,417   &   0.474      &  6.7$\times10^{-4}$   \\
\hline\hline
\end{tabular}
\label{data}
\end{table}
\begin{figure}[htbp]
\includegraphics[width=8.8cm,height=7.2cm]{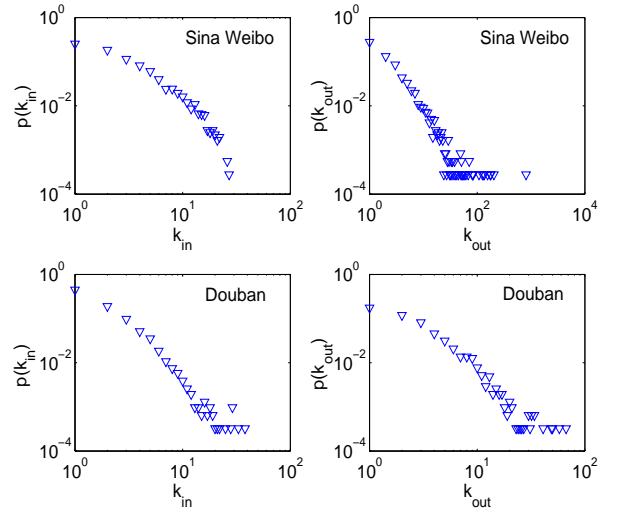}
\caption{In-degree (left) and out-degree (right) distributions of the two observed data sets. }
\label{data-degreedistribution}
\end{figure}

Consequently, such relationship can be represented by a directed network $G(V,E)$, where $V$ is the
set of nodes and $E$ is the set of edges. Each node represents a user, and one link from user $i$ to
user $j$ indicates $i$ is followed by $j$, that is to say,
$i$ is the \emph{followee} of $j$, and $j$ is one of $i$'s \emph{follower}. Table~\ref{data} summarizes the basic
statistics of the observed datasets. In addition, Fig.~\ref{data-degreedistribution} shows the out-degree distributions
which power-law $p(k_{out})\propto k^{-\lambda}_{out}$ with exponents $\lambda$=1.366 and 1.958,
for \emph{Sina Weibo} and \emph{Douban}, respectively. This common feature suggests that most users
are ordinary beings who have relative small number of followers and keep only a small fraction of celebrities.
Comparatively, the in-degree distribution of the two datasets does not exhibit the same phenomenon.
The in-degree distribution of \emph{Douban} still keep power-law shape with exponent 2.387,
but \emph{Sina Weibo} has a cut-off around $k_{in}=20$. One possible reason is that \emph{Sina Weibo} only allows
a certain number of followers for each free account. It might also suggest the different mechanisms driving the growth
of two sites: information diffusing automatically in microblogging system of \emph{Sina Weibo},
comparing with the information filtering by recommendation-related technique in \emph{Douban}. Similar difference between passive and automatic patterns was also empirically reported in bipartite and hypergraph networks~\cite{Shang2010, ZhangZK2011}.
In addition, we further investigate the average number of common follower and followees (see Table~\ref{common-neighbor}). Compared to non-reciprocal node pairs, reciprocal ones tend to have more common followers and followees, which is in accordance with previous work~\cite{Cui2012}.

\begin{table}[htbp]
\caption{Comparisons of the average number of common followees ($N_{CI}$) and followers ($N_{CF}$) for reciprocal and non-reciprocal node pairs, respectively.}
\begin{tabular}{C{2.5cm}|C{1.0cm}C{1.0cm}C{1.0cm}C{1.0cm}}
\hline\hline &  \multicolumn{2}{c}{Sina Weibo}  & \multicolumn{2}{c}{Douban} \\ \cline{2-5}
  & $N_{CI}$ & $N_{CF}$ & $N_{CI}$ & $N_{CF}$  \\
\hline
Reciprocal & \textbf{0.498} & \textbf{0.451}&   \textbf{0.215} & \textbf{0.155}\\
Non-reciprocal & 0.274 & 0.295 &0.029 & 0.083  \\
\hline\hline
\end{tabular}
\label{common-neighbor}
\end{table}

\section{Methods and Results} To better understand the influence of link reciprocity in social networks,
in the following, we shall evaluate its effects on information spreading and network robustness from the
perspectives of the network function and structure, respectively.

\subsection{Effect on Information Spreading}
Information spreading~\cite{Vespignani2011} is one of the most important functions of social networks, where the information (messages, tweets, comments, etc.) can distribute at a remarkably fast speed through the whole online society via frequent interactions among users, although its structure is not designed on purpose for spreading news~\cite{Doer2012}. Up to now, there is a considerable number of theoretical models to study information diffusion on social networks ~\cite{liuzonghua-2003PRE,Moreno-2004PRE,Hill-2010Plos,House-2011,Dodds-2004PRL}. In this Letter, in order to understand the underlying mechanisms and possible factors that would result in the information outbreaks, we adopt the classic epidemic spreading model, \emph{Susceptive-Infected} (SI) model \cite{Pastor2001}, to evaluate the effect of reciprocal links in the two aforementioned social networks. The diffusion process is described as following,

\begin{itemize}
\item Initially, user $i$ publishes an information item, \emph{I}, in the corresponding social network. \emph{I} could be about a piece of news, a photo, a comment, etc;
\item All $i$'s followers will automatically receive \emph{I} according to the \emph{follower-followee} directed network structure. Then an arbitrary fraction  of those followers might notice \emph{I}, and forward it on their own homepages if they find it interesting. We consider this \emph{forwarding willingness} as the \emph{transmission probability}, denoted by $p$;
\item The above step will be repeated to the followers of $i$'s followers, and eventually diffuses to the all achievable network nodes.
\end{itemize}

\begin{figure*}[htbp]
\centering
\includegraphics[width=18cm,height=10cm]{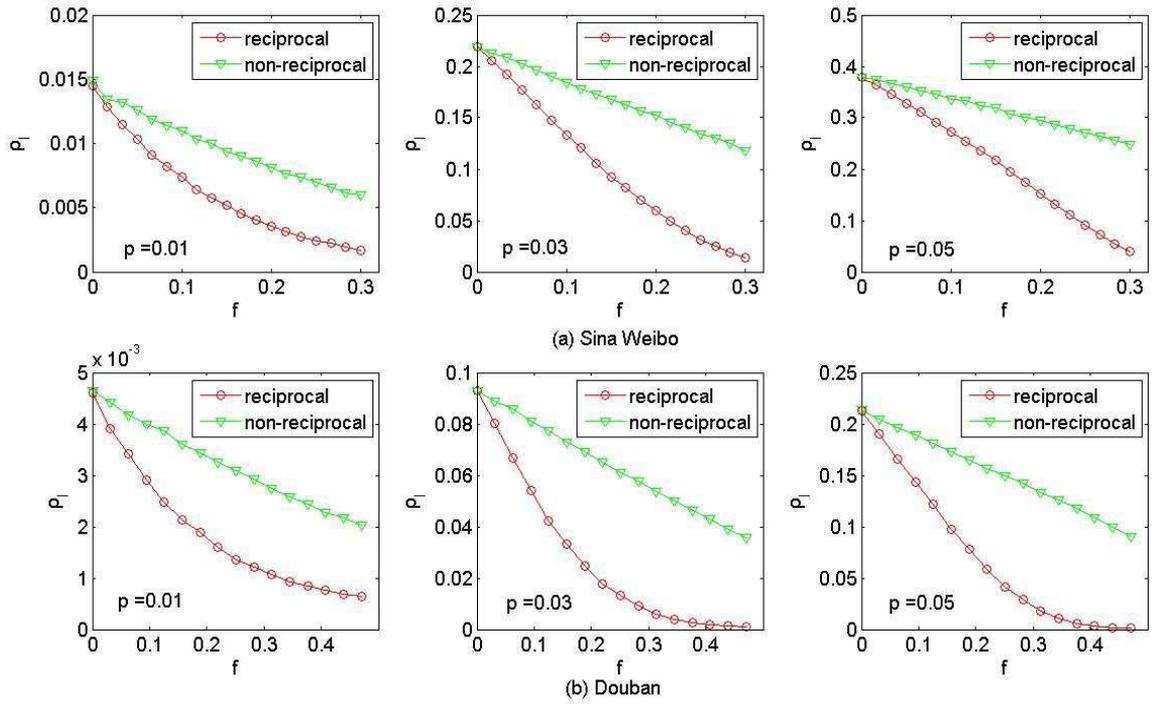}
\caption{(Color online) The fraction of influenced nodes as the function of the fraction of removed links $f$. In each subgraph, the red and green curves
correspond to removing reciprocal and non-reciprocal links, respectively. The experimental results are averaged over 30 independent realizations.}
\label{data-spreadspan}
\end{figure*}

\begin{figure*}[htbp]
\centering
\includegraphics[width=18cm,height=10cm]{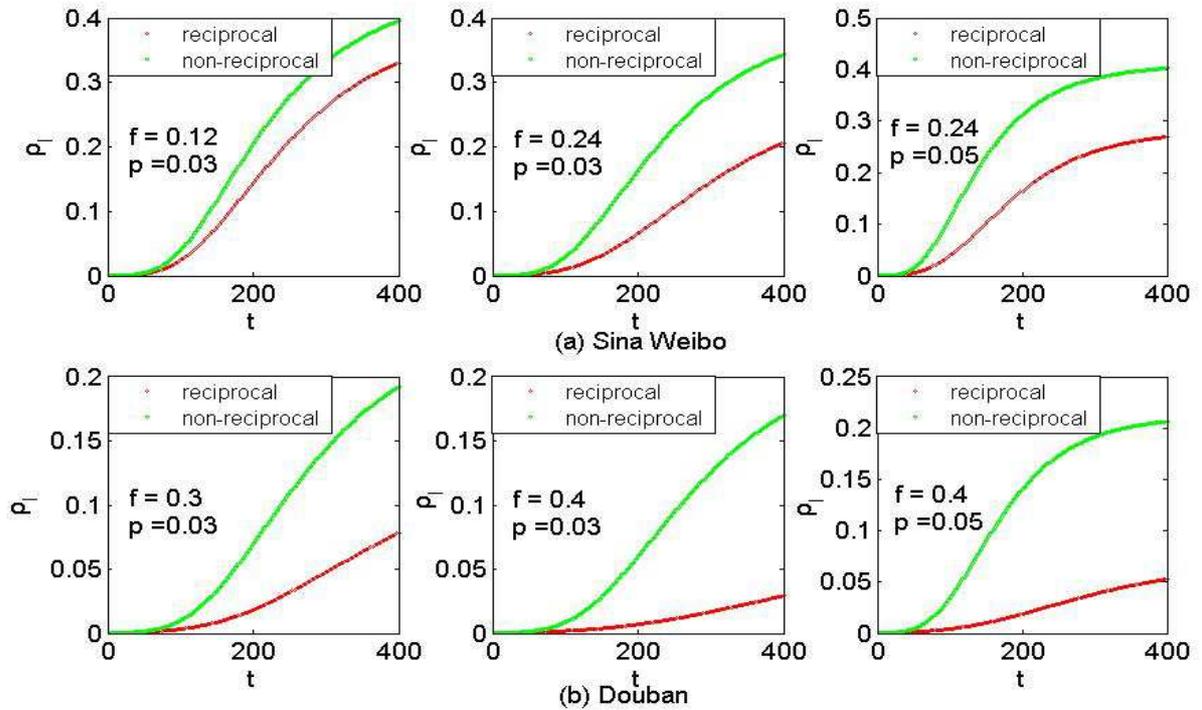}
\caption{(Color online) The fraction of influenced nodes as the function of observed time-step $t$, where $f$ is the fraction of removed
links. The red and green curves correspond to removing reciprocal links and non-reciprocal links, respectively. The experimental results are averaged over 30 independent realizations.}
\label{data-spreadspeed}
\end{figure*}

Note that, the main difference between the directed \emph{SI (DSI)} and classical \emph{SI} model is that the link direction is taken into account. In the proposed \emph{DSI} model, the information only can be transmitted from the followee to its own followers along with the direction of edges. Therefore, the final fraction of influenced nodes, $\rho_{I}$, is determined by such a structure. In order to observe the effects of reciprocal links on information diffusion, we quantify the influence according to an edge percolation process ~\cite{Onnela, Juan-2008PRL, Schwartz-2002PRE, ChengXQ2010JSM}. Obviously, if one reciprocal link is more important than two separate non-reciprocal links, the information diffusion results will be affected significantly when we remove the same fraction of reciprocal and non-reciprocal links. Fig.~\ref{data-spreadspan} compares the information coverage of removing the two types of links. Compared with removing non-reciprocal links, $\rho_{I}$ decays much faster when we remove the same amount of reciprocal links. Analogously, it also can be seen from Fig.~\ref{data-spreadspeed} that the diffusion speed is affected much remarkably when removing reciprocal links. Therefore, it demonstrates that reciprocal links indeed play a more important role in the information diffusion process on directed social networks.

\subsection{Effect on Structural Robustness}
In conventional complex network theory, it is wildly agreed that the network function is largely influenced by its specific structure \cite{Newman200301}. Therefore, to give solid and comprehensive understanding of the aforementioned results, we adopt the a dynamical removing process to measure the effects of reciprocal links on maintaining the structural robustness of networks~\cite{ChengXQ2010JSM}. For comparison, we apply three metrics to quantify the corresponding performance. (i) $R_{GSCC}$: the relative size of the strongly connected giant component. A sudden decline of $R_{GSCC}$ will be observed if the network disintegrates
after deleting a certain fraction of edges; (ii) the network susceptibility ($\tilde{S}$): defined as
\begin{equation}
 \tilde{S} = \sum\limits_{s<s_{max}}\frac{n_{s}s^{2}}{N},
\label{equation-S}
\end{equation}
where $n$ is the number of components with $s$ nodes, $N$ is the size of the network, and the sum runs
over all the components except the largest one ($s_{max}$). Note that, different with the definition in undirected networks, in Eq. (\ref{equation-S}), we only
consider the strongly connected component in directed networks. Considering $\tilde{S}$ as the function of the fraction of removed
edges $f$, usually, an obvious peak can be observed when the network disintegrates~\cite{Stauffer-1994, Bunde-1996};
(iii) the average distance $\langle d \rangle $, calculated by
\begin{equation}
 \langle d \rangle = \frac{1}{N(N-1)}\sum\limits_{<i,j>\in E,i\neq j}d_{<i,j>},
\label{equation-d}
\end{equation}\\
where $d_{<i,j>}$ is the distance from node \emph{i} to \emph{j}. $d_{<i,j>}$ is set to $N$ when there is no directed path
from node \emph{i} to \emph{j}. Clearly, the smaller $\langle d \rangle$ is, the better connectivity and more efficient the network will be.

\begin{figure}[htbp]
\includegraphics[width=8.8cm,height=7cm]{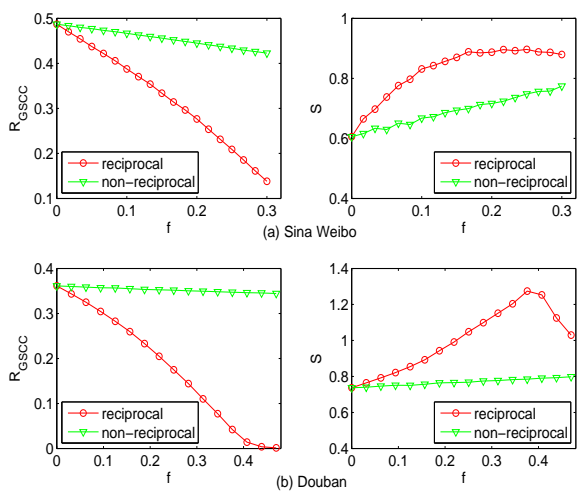}
\caption{(Color online) The fraction of giant component size ($R_{GSCC}$) and the susceptibility ($\tilde{S}$) as the function of the fraction of removed links $f$ on the two observed datasets, (a) \emph{Sina Weibo} and (b) \emph{Douban}. In each subgraph, the red and green curves correspond to the results of reciprocal and non-reciprocal links, respectively. The experimental results are averaged over 30 independent realizations.}
\label{data-gscc}
\end{figure}

\begin{figure}[htbp]
\includegraphics[width=8.8cm,height=3.3cm]{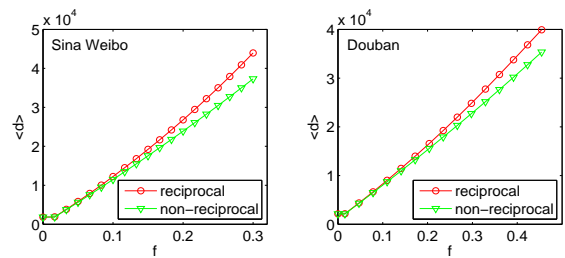}
\caption{(Color online) The average network distance ($\langle d \rangle$) as the function of removed links $f$ on the two observed datasets, (left panel) \emph{Sina Weibo} and (right panel) \emph{Douban}. The red and green curves correspond to the results of reciprocal and non-reciprocal links, respectively. The experimental results are averaged over 30 independent realizations.}
\label{data-pathlength}
\end{figure}

Fig.~\ref{data-gscc} and Fig.~\ref{data-pathlength} show the corresponding results of the three examined matrices. In Fig.~\ref{data-gscc}, it shows different dynamical patterns of removing reciprocal and nonreciprocal links, respectively. The size of strongly connected giant component ($R_{GSCC}$) decreases more sharply when removing reciprocal links than deleting non-reciprocal ones. Accordingly, the network susceptibility ($\tilde{S}$) result shows a percolation phenomenon when removing reciprocal links. Comparatively, this phenomenon is not observed when removing non-reciprocal links. In addition, Fig.~\ref{data-pathlength} shows that the average network distance ($\langle d \rangle$) increases much faster when removing reciprocal links than deleting the nonreciprocal ones. In a word, different dynamical results indicate that reciprocal links play a more important role in both maintaining the connectivity and keeping the efficiency of directed networks than non-reciprocal links. It also strongly supports the results in the previous section that reciprocity can much promote the speed of information diffusion, as it takes a more significant responsibility for the robustness of directed networks.

\section{Conclusions and Discussion}
In this Letter, we have studied the influence of reciprocal links of directed networks from two perspectives: (i) information spreading; (ii) structural robustness. Experimental results on two representative directed social networks, \emph{Sina Weibo} and \emph{Douban}, show that reciprocal links indeed play a more important role than non-reciprocal ones. In particular, the results of information spreading show that reciprocity can significantly enhance the spreading speed. In addition, the corresponding observations on the two examined datasets show that the reciprocity is also largely responsible for maintaining the connectivity and keeping the efficiency of directed networks, which suggests its significant impact in information spreading on networks.

The findings of this work may have a wide-range application in studying the role and influence of reciprocal links. Firstly, the topic of community detection has been well discussed \cite{Fortunato2010}, however, the progress on directed networks \cite{Leicht2008PRL} is relatively slow. The main reason is that the modularity \cite{Girvan2002PNAC} of directed networks is rather difficult to be precisely defined. Secondly, most studies on epidemic spreading and information diffusion \cite{Lv2011NJP} focus on studying the corresponding dynamics on undirected networks, the in-depth theoretical understanding of the underlying spreading mechanism on directed networks \cite{LiuC2012} still remains to be solved. Finally, the area of information filtering \cite{Lv2012} confronts a huge challenge as more and more directed social services are provided in the information era. The present work just provides a start point to see the preliminary effects of reciprocal links, a more comprehensive and in-depth understanding of reciprocity still need further efforts to discover.


\acknowledgments This work was partially supported by the National Natural Science Foundation of China
(Grant Nos. 11105024, 11105025, 1147015 and 11205040). ZKZ acknowledges the Zhejiang Provincial Natural Science Foundation of China (Grant Nos. LY12A05003 and LQ13F030015), the start-up foundation and Pandeng project of Hangzhou Normal University. ZYX acknowledges the Fundamental Research Funds for Central Universities (Grant No. A03008023401042).

\end{document}